# Near-field scanning microwave microscope for interline capacitance characterization of nanoelectronics interconnect

Vladimir V. Talanov and Andrew R. Schwartz

*Abstract—* **We have developed a noncontact method for measurement of the interline capacitance in Cu/low-*k* interconnect. It is based on a miniature test vehicle with net capacitance of a few femto-Farads formed by two 20-μm-long parallel wires (lines) with widths and spacings the same as those of the interconnect wires of interest. Each line is connected to a small test pad. The vehicle impedance is measured at 4 GHz by a near-field microwave probe with 10 μm probe size via capacitive coupling of the probe to the vehicle's test pads. Full 3D finite element modeling at 4 GHz confirms that the microwave radiation is concentrated between the two wires forming the vehicle. An analytical lumped element model and a short/open calibration approach have been proposed to extract the interline capacitance value from the measured data. We have validated the technique on several test vehicles made with copper and low-*k* dielectric on a 300 mm wafer. The vehicles interline spacing ranges from 0.09 to 1 μm and a copper line width is 0.15 μm. This is the first time a near-field scanning microwave microscope has been applied to measure the lumped element impedance of a test vehicle.**

*Index Terms—* **Interconnect, interline capacitance, low-*k* dielectric, near-field scanning microwave microscopy, lumped element impedance.**

## I. INTRODUCTION

INTERLINE capacitance is an important interconnect metric often governing the overall operating performance of today's advanced integrated circuits (ICs). Along with the wire resistance it affects such critical interconnect parameters as signal delay time (RC time constant), crosstalk, and power consumption [1]. The interline capacitance also carries important information about the interconnect integration in which the copper wiring is inlaid in the trenches and vias pre-etched in the interlevel dielectric. Such information could be associated with, for example, the manufacturing tolerances or increase in the dielectric constant due to plasma damage induced in porous low-*k* dielectrics by various processing steps [2]. Therefore, acquiring the interline capacitance non-destructively and analyzing it in a timely manner can benefit both the development and production stages of interconnect integration, while addressing the semiconductor industry's need for *in-line* and *in-situ* metrologies [3].

At present, the only technique capable of measuring interconnect interline capacitance is the interdigital comb capacitor that is formed by two identical comb-shaped structures interleaved in such way that the fingers alternate [4] (such geometry is often employed in microwave circuitry to create a lumped element capacitor). The measurement is typically done with an LCR meter at 0.1-1 MHz via contacting the test pads connected to the combs. To overcome the effect of parasitic capacitances associated with the pads and LCR leads the comb capacitance has to be greater than a few pico-Farads, which requires the structures to be larger than 100 × 100 μm in size. The comb accuracy is typically limited to 10 to 50 % due to stray fields, fringing effects, and statistical pattern variation across the comb [1, 5]. Usually multiple interdigital capacitors are patterned onto the wafer to address a wide range of interline spacings and/or line widths. While the comb capacitor is routinely implemented on patterned wafers during new process development and represents real interconnect properties, it cannot be used for *in-line* metrology because it consumes significant real estate while extensive sample cross-sectioning and numerical modeling are needed to analyze the measurements.

Our goal has been to develop a technique to measure the interline capacitance on semiconductor production wafers that overcomes the above shortcomings. The desired test vehicle needs to be small enough to fit into the 80-μm-wide wafer scribe-line (i.e., the space between active die regions on the wafer), be compatible with standard interconnect processing flow, and be repeatable at all interconnect levels. The measurement must be noncontact, noninvasive, and provide for real time data collection and analysis. To address these requirements we designed a miniature femto-Farad test vehicle and measured its impedance by a near-field scanning microwave microscope at 4 GHz.

The spatial resolution in near-field (evanescent) scanning probe microscopy is governed by the probe size rather than the freespace wavelength *λ* of the radiation [6]. Unlike conventional far-field diffraction-limited microscopy where the observation of sub-wavelength structures is limited because the smallest spot to which the light can be converged

V. V. Talanov is with Solid State Measurements, Inc., 110 Technology Drive, Pittsburgh, PA 15275, USA (phone: 412-787-0620; fax: 412-787-0630; e-mail: vtalanov@ssm-inc.com).

A. R. Schwartz was with Solid State Measurements, Inc. He is now with the US Department of Energy, Office of Basic Energy Sciences (e-mail: Andrew.Schwartz@science.doe.gov).

is about $\lambda/2$ due to the Abbe diffraction limit, the near-field can be confined to a region that is orders of magnitude smaller than the wavelength. This is possible because the electrodynamic response of a near-field microwave probe is due to reactive energy, electric and/or magnetic, stored in the near-field (or evanescent waves or near-zone field) in the vicinity of the probe. Because of its static nature the near-field is confined to a volume governed by the probe size and therefore the microscope's spatial resolution is defined by the probe geometry rather than the wavelength. Near-field scanning probe microwave microscopes with spatial resolution down to $\lambda/10^6$ have been demonstrated (see, for example, [6-8] and references therein). Near-field coupling of a probe to a sensor or of a transmitter to a receiver antenna has been explored in the biomedical and wireless fields [9, 10].

The near-field microwave microscope apparatus employed in this paper provides a moderate spatial resolution of about 10 μm (or $\lambda/10^4$). Such a probe size is very well suited for precise and accurate dielectric constant measurements of 100 to 1000 nm thick films [11]. It also appears to be optimal for blanket low-$k$ dielectric constant measurements on semiconductor production wafers [12] using test vehicles of about 50 × 50 μm$^2$ in size. In the present paper we demonstrate how the probe formed by a balanced parallel strip transmission line, unique in near-field scanning microwave microscopy, can be used for a noncontact measurement of a test vehicle's lumped element impedance (preliminary results were presented in [13]). While in the past decade near-field microwave microscopes have been extensively employed for nondestructive imaging of materials' linear and non-linear electrodynamic properties as well as electromagnetic fields in microwave circuits [6-8], a measurement of the test structure impedance has not yet been demonstrated. Some advantages of using a microwave microscope versus an on-wafer probe for such an application are noncontact operation and higher capacitance sensitivity due to reduced parasitic capacitances associated with the vehicle contact pads and LCR leads.

## II. EXPERIMENTAL METHODS AND THEORY

### A. Near-field scanning probe microwave microscope

Our apparatus has been described in detail elsewhere [14]. The probe is based on a 25-mm-long tapered quartz bar sandwiched between two 2-μm-thick aluminum strips. The bar cross-section reduces from 1.2 × 1 mm at the probe upper end to approximately 6 × 5 μm at the probe tip (see Fig. 1) with aluminum covering the two wider sides of the bar. This forms a tapered broad-side coupled parallel strip transmission line that can carry a balanced quasi-TEM odd mode with an effective dielectric constant of 2.3. A strong and highly confined electrical sampling field (similar to the fringe field of a parallel plate capacitor) is created between the two aluminum electrodes at the electrically open tip facet, which is micromachined by a focused ion beam technique similar to [15]. We would like to point out here that a waveguide of similar geometry has been proposed by Klein et al. [16] for terahertz near-field imaging.

To improve the measurement sensitivity the parallel strip transmission line is formed into a half-lambda resonator with a fundamental resonant frequency of 4 GHz and an unloaded $Q$-factor of about 100. A magnetic coupling loop is employed to excite the resonator. The loop position is adjusted to yield critical coupling to the resonator, thereby matching the impedance of approximately 1 MΩ terminating the probe tip to the 50 Ω characteristic impedance of the coaxial feed line [17]. The resonator and the coupling loop are packaged into a cylindrical aluminum enclosure with the resonator tapered end protruding a few millimeters out via a clear hole in the enclosure end wall. The probe frequency is measured with better than 1 kHz sensitivity by a frequency counter tapped into a frequency tracking loop similar to [18].

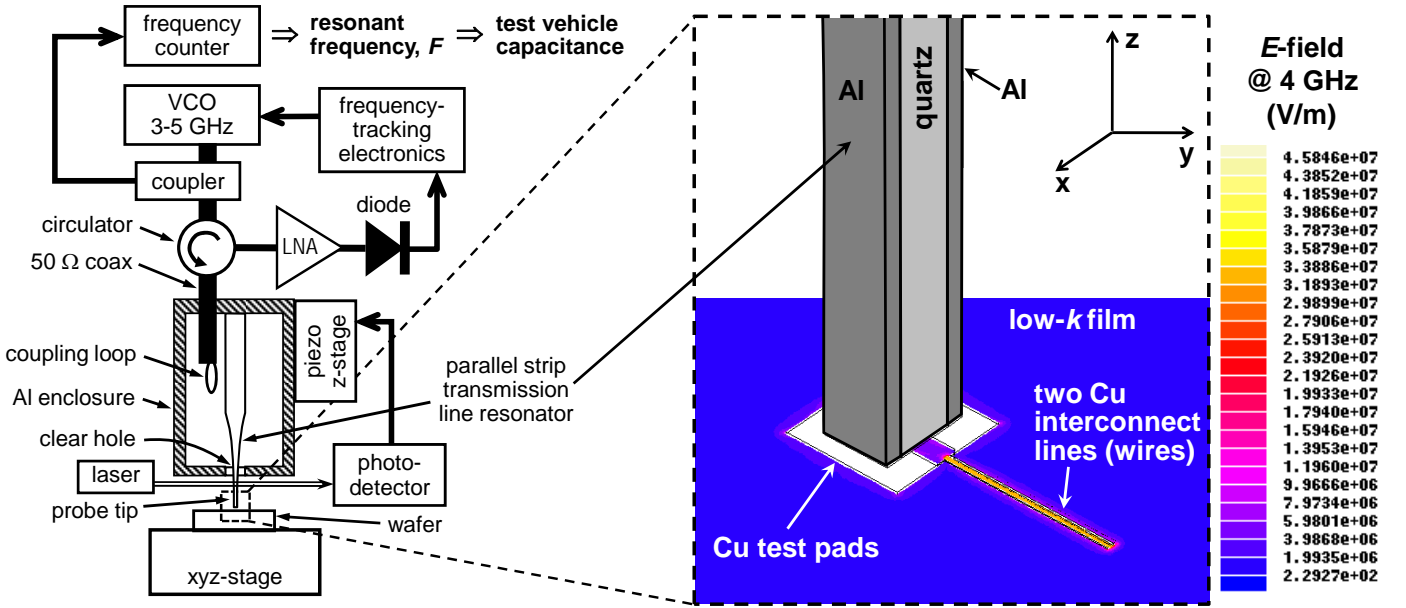

Fig. 1. Schematic of the near-field scanning microwave microscope apparatus (VCO is the voltage controlled oscillator, LNA is the low noise amplifier). Inset shows a zoom-in of the probe tip capacitively coupled to the test vehicle (that is also shown in Fig. 2a). The intensity plot is the electric field in xy-plane located 0.15 μm below the sample surface, as obtained with full 3D finite element modeling at 4 GHz. The probe tip facet size is 6 × 9 μm, the tip-sample air-gap is 0.1 μm, the separation between the two copper interconnect lines (interline spacing) is 0.2 μm, and the interconnect lines length is 20 μm. Note a strong electric field concentration between the two interconnect lines (also see Fig. 3). The fields in the parallel strip transmission line are not shown for clarity.

The probe tip to sample distance $h$ is maintained at less than 100 nm using a shear-force signal feedback idea originally proposed in near-field scanning optical microscopy [19, 20]. The tapered quartz bar forms a cantilevered beam in which a fundamental vibration mode at 2.6 kHz is excited by dithering the enclosure with nanometer amplitude using a piezo tube. The probe tip is illuminated with a laser beam projecting onto a photo-detector (see Fig. 1) and the ac output of the detector depends on the tip vibration amplitude, which is a strong function of the tip-sample distance for $h < 100$ nm. This signal is fed into a closed loop control circuit that drives a piezo z-stage capable of moving the probe enclosure up and down with sub-nanometer resolution. To eliminate drift in the tip-sample air-gap the set point of the shear-force control loop is periodically recalculated.

One can employ the probe resonant frequency $F$ to estimate the actual precision of the distance control feedback. This is done by measuring the microwave resonant frequency standard deviation $\sigma_F$ on a sample, such as a low resistivity Si wafer, where the *apparent* $\sigma_F$ (typically 20-50 kHz depending on the actual probe) is much greater than the microscope's frequency sensitivity. The standard deviation $\sigma_h$ for the tip-sample distance $h$ is then calculated as $\sigma_h = \sigma_F \left( \partial F / \partial h \right)^{-1}$, where the derivative $\partial F / \partial h \sim 100$ kHz/nm is experimentally determined by measuring $F$ vs. $h$ just above the shear-force distance. Using this procedure we found $\sigma_h$ to be about 1 nm.

The apparatus is equipped with an optical vision system to locate and navigate to patterned structures. It sits on a vibration-isolated platform inside an environmental chamber at ambient conditions. Wafers up to 300 mm in diameter can be scanned beneath the probe.

### B. Femto-Farad test vehicle

The proposed test vehicle for the interline capacitance metrology is shown in Fig. 2(a). It is formed by two parallel copper lines with width and spacing that are the same as those of interconnect wires of interest. The lines are connected to the test pads to allow the near-field probe be capacitively coupled to the vehicle. The vehicle is embedded into a low-*k* dielectric and made flush with its surface by chemical mechanical planarization (see Fig. 3). Such a vehicle can also be viewed as an interdigital capacitor with only two fingers.

Fig. 1, inset shows the result of full 3D finite element modeling performed with Ansoft's High Frequency Structure Simulator (HFSS) [21]. The tip end portion of a parallel strip transmission line was driven in the fundamental mode at 4 GHz. The electric field intensity is shown in the plane located 0.15 μm below the sample surface and parallel to the *xy*-plane.

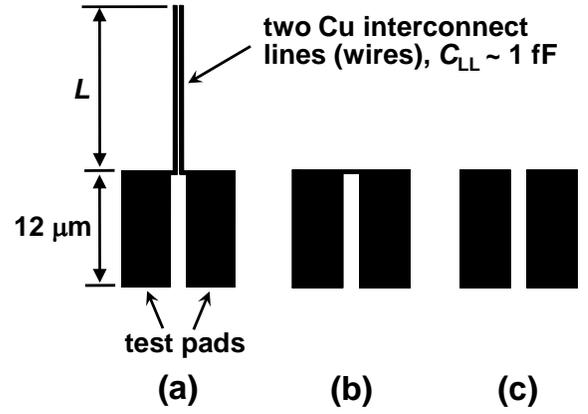

Fig. 2. In-plane view of **(a)** the interline capacitance test vehicle with $L$ ranging from 10 to 40 μm, **(b)** "short", and **(c)** "open" calibration test keys. The net capacitance between the two interconnect lines $C_{LL}$ is on the order of a few femto-Farads.

As expected, the microwave radiation is efficiently coupled into the vehicle and the electric field is concentrated in the region between the two interconnect lines. To confirm this we studied the electric field distribution in a plane parallel to the *xz*-plane, which intersects the two interconnect lines from Fig. 1, inset. The result is shown in Fig. 3. Note that the simulation quality is moderate here because the span of dimensional features in Fig.1, inset (e.g., from 0.1 μm for the tip-sample air-gap to about 100 μm for the entire simulation box) was near the limit of the HFSS engine.

### C. Test vehicle capacitance

Because both the tip and the test vehicle are much smaller than the 7.5 cm free space wavelength there is no phase variation across the vehicle and a lumped element description is valid. The net capacitance of the near-field probe tip capacitively coupled to the test vehicle (see Fig. 1, inset) is

$$C_{tv} = \left( \frac{1}{C_C} + \frac{1}{C_{LL} + 0.5 C_{PG} + C_{PP}} \right)^{-1} \tag{1}$$

where $C_C$ is the tip to test pads total coupling capacitance, $C_{LL}$

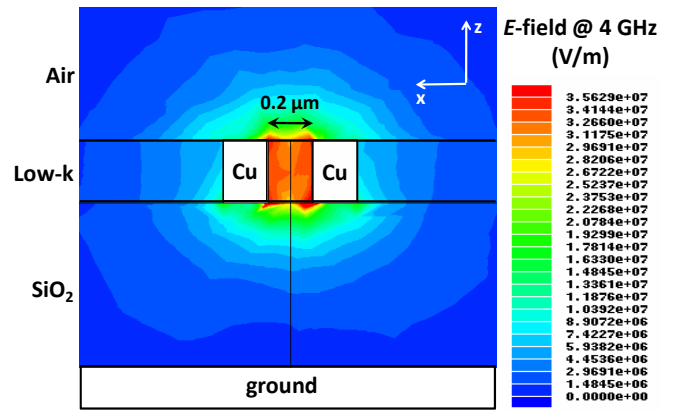

Fig. 3. Intensity plot of electric field exited at 4 GHz by the parallel strip transmission line probe between the two copper interconnect lines from the same simulation as in Fig.1, inset. Electric field is shown in the *xz*-plane plane cross-sectioning the two copper interconnect lines.

is the net interline capacitance between the two interconnect lines under test, $C_{PG}$ is the test pad to ground (or pad to underlying layers) capacitance, and $C_{PP}$ is the capacitance between the test pads. Since $C_{PG}$ and $C_{PP}$ are connected in parallel to $C_{LL}$ they are parasitic.

To eliminate the unknown quantities $C_C$, $C_{PG}$ and $C_{PP}$ from (1) we designed the "short" and "open" calibration keys shown in Figs. 2b and 2c, respectively. The net tip capacitances for the probe coupled to the "short" and "open" calibration keys are

$$C_{sh} = C_C \tag{2a}$$

and

$$C_{op} = \left( \frac{1}{C_C} + \frac{1}{0.5C_{PG} + C_{PP}} \right)^{-1}, \tag{2b}$$

respectively. Thus (1) and (2) yield the following expression for the capacitance between the two interconnect lines $C_{LL}$

$$C_{LL} = \frac{C_{sh}^2(C_{tv} - C_{op})}{(C_{sh} - C_{op})(C_{sh} - C_{tv})} \tag{3}$$

The relationship between $C_{LL}$ and the specific interline capacitance (i.e., capacitance per unit length) $C_{sp}$ is $C_{LL}=C_{sp}L(1+\alpha S/L)$, where $S$ is the interline spacing, and $\alpha S/L$ accounts for the fringe effect contribution at the end of the interconnect lines. It can be shown that parameter $\alpha$ is on the order of unity. Thus, for typical wire geometry, such as $L$=20 μm and $S$=0.2 μm, the fringe effect contribution is on the order of 1%. We used $\alpha$=1 for all of our calculations.

### D. Probe-vehicle interaction

When the probe tip is brought in close proximity and capacitively coupled to a test vehicle which impedance is substantially capacitive (see Fig.1, inset), the energy stored in the parallel strip transmission line resonator reduces and the resonant frequency $F$ decreases. The complex resonant condition (e.g., see [22]) applied to the resonator yields the following relationship between the net impedance $Z_t=R_t+iX_t$ terminating the probe tip and the relative change in $F$

$$\frac{\Delta F}{F_0} = \frac{Z_0}{\pi} \Delta \left( \frac{X_t}{R_t^2 + X_t^2} \right) \tag{4}$$

where $Z_0$ is the characteristic impedance of the parallel strip transmission line, $F_0$ is the probe resonant frequency with no sample present, and the second and high order terms associated with $|Z_t| \gg Z_0$ and $Q \gg 1$ are neglected. For typical interconnect geometries our test vehicles provide $R_t$ and $|X_t|$ on the order of 10 Ω and 1 MΩ, respectively. Therefore, we can

substitute $R_t \ll X_t$ and $X_t = -1/\omega C_t$ ($\omega$=2πF) into (4) which yields:

$$\frac{\Delta F}{F_0} = -2F_0 Z_0 \Delta C_t \tag{5}$$

where $C_t$ is the net near-field probe tip capacitance, and $(2FZ_0)^{-1} = 1.25$ pF is the resonator capacitance. For a typical frequency measurement sensitivity of 1 kHz (5) shows that the microscope's sensitivity to $C_t$ is about 0.3 aF. This allows for a test vehicle with capacitance on the order of 1 fF, and therefore significantly reduced size as compared to typical interdigital capacitors. For comparison, state of the art capacitance sensitivity of less than a zepto-Farad has been demonstrated in scanning capacitance microscopy at several GHz [23].

Finally, from (3) and (5) we relate $C_{LL}$ to the probe resonant frequency as follows

$$C_{LL} = \frac{(F_0 - F_{sh})^2}{2Z_0 F_0^2} \frac{(F_{op} - F_{tv})}{(F_{op} - F_{sh})(F_{tv} - F_{sh})} \tag{6}$$

Here $F_{tv}$, $F_{sh}$, and $F_{op}$ are the probe resonant frequencies for the test vehicle, "short" calibration key, and "open" calibration key, respectively. To derive (6) we neglected $2F_0C_0Z_0$ in comparison to unity, where $C_0 < 0.1$ fF is the tip capacitance with no sample present. All the frequency associated terms in the right hand side of (6) are determined with accuracy better than 1%. Thus the accuracy of $C_{LL}$ is mostly limited by the knowledge of the characteristic impedance $Z_0 \approx 100$ Ω that was determined numerically using HFSS. Given the practical uncertainties in the parallel strip transmission line geometry and dielectric constant one may expect the $Z_0$ accuracy to be better than 10%.

### E. Optimization of test vehicle geometry

Because of the limitations imposed by the probe tip, the shear-force distance, and the sample on the range of values for $C_C$, $C_{PP}$ and $C_{PG}$ (and therefore the calibration quantities $C_{sh}$ and $C_{op}$) the only parameter which can be varied to optimize the method's sensitivity to the specific interline capacitance $C_{sp}$ is the interconnect line length $L$. To find an optimum $L$ value we used the error propagation analysis (e.g., see [24]) to calculate a standard deviation for the specific interline capacitance $\sigma_{Csp}$. The two experimentally measured variables are the probe frequency $F$ and the tip-sample distance $h$, for which the covariance term was assumed to be zero. The following parameters were used: frequency measurement standard deviation $\sigma_F$=1 kHz, tip-sample distance standard deviation $\sigma_h$=1 nm, tip-sample shear-force distance $h$=80 nm (see [9]), pad size 12 × 3 μm, and low-$k$ stack parameters as described in the next section. Fig. 4 shows a semi-logarithmic plot of $\sigma_{Csp}/C_{sp}$ versus $L$ for several specific interline capacitances in the range of interest. At small $L$ the method sensitivity is mostly limited by the parasitic capacitances $C_{PG}$ and $C_{PP}$, as can also be seen from (1). At large $L$ where $C_{LL}$ is greater than $C_{PP}$ and $C_{PG}$ the sensitivity is limited by the shear-force distance control (i.e., by $C_C$) repeatability. One can see that the optimum line length varies substantially with $C_{sp}$. We have chosen $L$ to be 20 µm which yields the sensitivity between 12 and 14 % for the entire range of interline spacings under test here. The error propagation analysis also shows that reducing $\sigma_h$ from 1 nm to 0.2 nm, which we recently observed in the state of the art experiments [25], would improve the method sensitivity to below 3%.

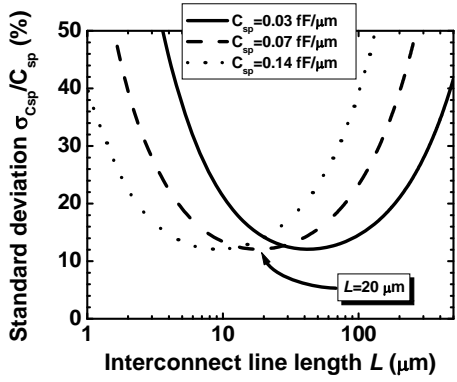

Fig. 4. Standard deviation of specific interline capacitance vs. the interconnect line length $L$ for $C_{sp} = 0.03$ (solid line), 0.07 (dashed line), and 0.14 (dotted line) fF/µm based on the error propagation analysis (see text).

## III. SAMPLE DESCRIPTION

To experimentally validate the proposed test vehicle design and calibration approach two sets of test vehicles were fabricated on a 300 mm wafer using a single damascene scheme. The stack information was as follows (bottom to top): Si/ 390 nm low-$k$ $k$=3.2/ 50 nm SiC $k$=4.2/ 360 nm low-$k$ $k$=3.2 (also see Fig. 7). The trench depth was 300 nm for all structures. Non-porous silica based low-$k$ material with nominal dielectric constant $k$=3.2 was chosen in order to minimize the potential dielectric damage during fabrication. The test vehicles in Set A had 20-µm-long copper lines, fixed 0.15 µm linewidth, and variable interline spacing $S$ of 0.09, 0.1, 0.12, 0.15, 0.2, 0.3, 0.6 and 1.0 µm. Set B consisted of four vehicles with 0.15 µm linewidth, fixed 0.3 µm interline spacing, and variable line length $L$ of 10, 20, 30 and 40 µm. Both sets also included the short and open calibration keys.

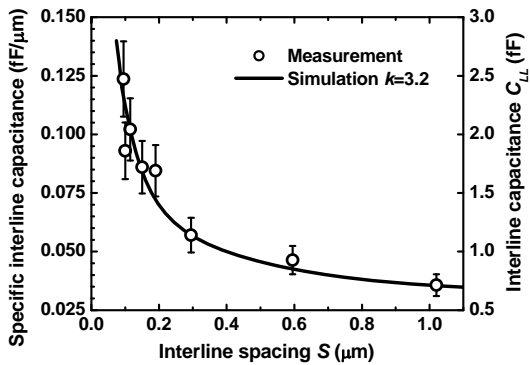

Fig. 5. Specific interline capacitance $C_{sp}$ and interline capacitance $C_{LL}$ vs. interline spacing $S$ for eight test vehicles with $L$=20 µm lines. Solid curve is the result of a 2D electrostatic simulation. The error bars are ± one standard deviation as estimated with the error propagation analysis (see text).

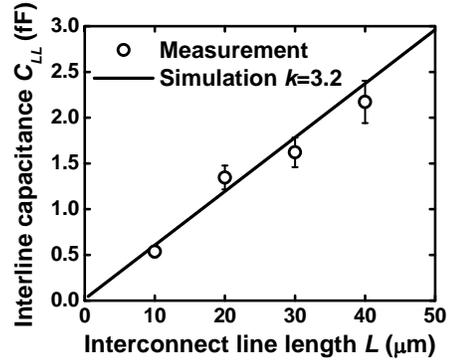

Fig. 6. Interline capacitance $C_{LL}$ vs. the interconnect line length $L$ for a set of four test vehicles with the same interline spacing $S$=0.3 µm. Solid line is the result of a 2D electrostatic simulation. The error bars are ± one standard deviation as estimated with the error propagation analysis (see text).

## IV. RESULTS AND DISCUSSION

Fig. 5 shows the results for Set A. $C_{sp}$ and $C_{LL}$ values extracted from the measurement using (6) are plotted versus the actual interline spacing $S$ obtained from the vehicle cross-sections imaged by a scanning electron microscope. The error bars are based on the error propagation model described above. The solid line is the result of 2D electrostatic finite element modeling, as shown in Fig. 7 (note that the distribution of the static field in Fig. 7 is very similar to that of the microwave field in Fig. 3, which confirms that the microwave energy is efficiently coupled into the vehicle). The experimental data and modeling results for Set B are shown on Fig. 6. Good quantitative agreement between the measurements and simulations is observed in both cases, while the deviation of data from a model is within the estimated method sensitivity.

## V. CONCLUSION

We have demonstrated the use of a near-field scanning probe microwave microscope for non-contact measurements of interconnect interline capacitance using a miniature femto-Farad test vehicle. Since the vehicle involves only two relatively short interconnect wires the measurement is nearly unaffected by the fringe effects and statistical pattern variation that can plague the conventional comb capacitor method. No electrical contact to, or grounding of, the wafer is required since both probe electrodes are located above and capacitively coupled to the test vehicle. Hence, the vehicle can be placed and/or repeated at any interconnect level. The proposed technique can be used to study the electrical parameter variations in IC interconnects as well as provide parameters for their statistical design. We are planning to investigate if it can also be employed to measure the resistance of interconnect lines as well as low-*k* dielectric losses. The developed test vehicle geometry has a potential to be employed for microwave characterization of various functional nano-materials such as nanowires, carbon nanotubes, nanofibers and nanoribbons by connecting the object under study between the two interconnect lines.

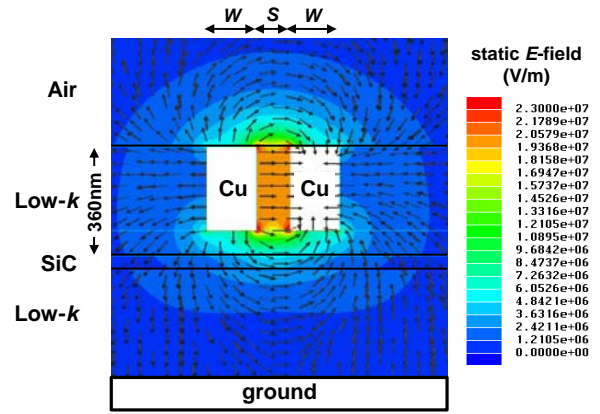

Fig. 7. 2D finite element electrostatic simulation of electric field between the two copper interconnect lines forming the test vehicle. $S$ is the interline spacing, and $W$ is the copper line width. Note a similarity of the field distribution with that in Fig. 3 which was obtained at 4 GHz.

### ACKNOWLEDGMENT

The authors are grateful to Dr. J. S. Tsai of TSMC for fabricating the wafer and Dr. A. Scherz for assistance with the measurements.